\documentclass[a4paper,11pt]{article}
\usepackage{pos}

\usepackage{enumitem}
\usepackage{caption}
\usepackage{subcaption}
\usepackage[
 print-unity-mantissa=false,
 exponent-product=\ensuremath{\cdot},
 uncertainty-mode = separate,
]{siunitx}
\title{The new architecture design of the Science Alert Generation pipeline of the Cherenkov Telescope Array Observatory}
\ShortTitle{The new architecture design of the CTAO SAG}
\renewcommand{\printHeadAuthors}{L. Castaldini et al.}

\author[a]{Luca Castaldini}
\author[a]{Andrea Bulgarelli}
\author[b]{Vincent Pollet}
\author*[a]{Gabriele Panebianco}
\author[b]{Pierre Aubert}
\author[b]{Sami Caroff}
\author[a]{Giovanni De Cesare}
\author[a]{Ambra Di Piano}
\author[a]{Valentina Fioretti}
\author[b]{Gilles Maurin}
\author[b]{Thibaut Oprinsen}
\author[a]{Nicolò Parmiggiani}
\author[b]{Thomas Vuillaume}
\author[c]{Igor Oya}
\author[d]{Kathrin Egberts}
\onbehalf{for the CTAO-ACADA Collaboration}

\affiliation[a]{INAF - Osservatorio di Astrofisica e Scienza dello spazio di Bologna, Via Piero Gobetti 93/3, 40129 Bologna, Italy}
\affiliation[b]{Univ. Savoie Mont Blanc, CNRS, Laboratoire d'Annecy de Physique des Particules - IN2P3, 74000 Annecy, France}
\affiliation[c]{CIEMAT, Avda. Complutense 40, 28040 Madrid, Spain}
\affiliation[d]{Institut für Physik und Astronomie, Univ. Potsdam, Germany}
\emailAdd{luca.castaldini@inaf.it}

\abstract{
The Cherenkov Telescope Array Observatory (CTAO) represents the next-generation gamma-ray observatory and will operate for several decades.
It will be particularly suited to analyse transients and variable phenomena, which will trigger real-time scientific alerts.
To support this, the Science Alert Generation (SAG) pipeline within the Array Control and Data Acquisition (ACADA) system will process data from telescope arrays in real time, using dedicated pipelines for data reconstruction (SAG-RECO), data quality monitoring (SAG-DQ) and science monitoring (SAG-SCI).
The Supervisor (SAG-SUP) oversees the dynamic operations of SAG and its integration with other ACADA components.

SAG is designed to issue candidate science alerts within $\SI{20}{\second}$ of data availability, processing events on multiple time scales (seconds to hours) and handling trigger rates of tens of kHz.
Meeting these requirements necessitates optimised software and hardware architectures.

This work presents recent developments in SAG's architecture, aimed at two main challenges: (1) selecting data only from telescopes that have entered a stable tracking state, even when they begin tracking at different times during multi-telescope observations, and (2) incorporating environmental and system monitoring information to ensure high data quality.
SAG-SUP can retrieve real-time telescope status and environmental conditions from telescope managers and the weather station through the ACADA Monitoring system, collect them in a database and then use them to filter out data from slewing phases or degraded conditions.

These enhancements are crucial to ensure the reliability of science alerts and improve the overall performance and responsiveness of the CTAO real-time analysis framework.
}

\ConferenceLogo{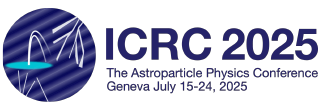}

\FullConference{39th International Cosmic Ray Conference (ICRC2025)\\
 15–24 July 2025\\
Geneva, Switzerland\\}

\begin{document}
\maketitle
\section{Introduction}
\label{sec:introduction}
The Cherenkov Telescope Array Observatory (CTAO) represents the next-generation facility for gamma-ray astronomy \cite{McMuldroch_CTAO_2024}.
The CTAO will provide extensive energy coverage from $\SI{20}{\giga\electronvolt}$ to $\SI{300}{\tera\electronvolt}$, allowing us to significantly advance our understanding of the Universe.
To maximise its scientific impact, the CTAO will issue scientific alerts, i.e. communications to the scientific community on the occurrence of transient and variable phenomena.
To do so, the CTAO requires a highly reliable and automated system to analyse the acquired data with low latency and issue candidate science alerts. A Candidate Science Alert is the notification that a transient event is detected, and it can require an immediate reschedule of an observation plan to perform the follow-up.
The Science Alert Generation (SAG) pipeline \cite{bulgarelli_SAGproceedingsSPIE_2024}, a key system of the Array Control and Data Acquisition (ACADA) system \cite{oya_acadaspie_2024}, fulfils this role by processing data from telescope arrays in real-time through its data reconstruction (SAG-RECO), data quality monitoring (SAG-DQ), science monitoring \cite[SAG-SCI,][]{Panebianco_SCI_ICRC2025} pipelines.
SAG also includes a Supervisor component (SAG-SUP), which is responsible for orchestrating operations and interfacing SAG with other ACADA sub-systems. The SAG pipeline will issue candidate science alerts to the Transients Handler system of ACADA.
SAG must meet strict functional requirements, including: (1) handling telescope trigger rates of tens of $\si{\kilo\hertz}$; (2) supporting multiple parallel analyses with different tools, configurations, and targets; (3) prioritising high-relevance scientific targets; (4) operating concurrently across different sub-arrays; and (5) being capable of issuing candidate science alerts with a maximum latency of $\SI{20}{\second}$ after the data becomes available. Meeting these demands requires dedicated, optimised software and hardware design.

A first version of SAG was delivered in ACADA Rel1 (2023), with a second version under development for ACADA Rel2 (expected 2025).
In this work, we present the latest updates to the architecture of the SAG-SUP system, under development during Rel2 and further releases to fulfil the functional requirements defined for the SAG system. These updates address two key challenges in the management of multi-telescope observations.
First, the system must select data only from telescopes that have reached a stable tracking state: the time required for a telescope to complete its slewing and begin stable tracking can vary significantly, from fractions of a second up to tens of seconds, depending on the telescope class (LST, MST, SST), repointing speed, and angular distance. This is particularly important for follow-up observations of transient sources, such as gamma-ray bursts (GRBs), where early data acquisition is crucial.
The system must include data from tracking telescopes without waiting for the full sub-array and exclude those still slewing, typically exhibiting larger reconstruction errors and degraded quality due to unstable pointing conditions.
Second, the system must account for varying observational conditions: adverse weather or hardware issues can degrade data quality, and the knowledge or estimation of such information must be used by SAG to select good-quality events.
We thus introduced a new component of SAG-SUP to discriminate which data must be processed according to telescope status and environmental conditions.

This work is organised as follows: in section~\ref{sec:SAGSUP}, we summarise the main interactions of SAG with ACADA and describe the architecture of SAG-SUP. 
We describe the enhancements to select data from tracking telescopes in section~\ref{sec:startDataTaking} and those for including monitoring information in section~\ref{sec:newGTI}. 
We discuss our results in section~\ref{sec:conclusion}.

\section{The Architecture and purpose of SAG-SUP in ACADA}
\label{sec:SAGSUP}
All ACADA sub-systems employ the Alma Common Software (ACS) as the basic middleware \cite{chiozzi}.
ACS is a framework for distributed systems that supports the development and integration of ACADA components and facilitates the operation of ACADA in the on-site data centre.
SAG's main interactions are with the Resource Manager (RM) which supervises all the other sub-systems; the Array Configuration System or Configuration Database (CDB) providing the deployment configuration; the Central Control (CC) responsible for the execution of the scheduled observations provided by the Short-Term Scheduler (STS); the Array Data Handler (ADH) \cite{Lyard_ADH_ICRC2025} which provides the input data from telescope cameras to the SAG; the Transient Handler (TH) \cite{Collins_TH_ICRC2025} to which SAG issues candidate scientific alerts; the Alarm system to which SAG sends alarms raised during the analyses; the Reporting system (REP) to generate reports; the Human-Machine Interface (HMI) to show results to the operators and support astronomers located in the control room of the CTAO.
In addition, SAG receives real-time environmental and night sky background level information from the Monitoring and Logging Systems (MON) \cite{Munari_AASMON_ICRC2025}.
As part of ACADA, the SAG pipeline also interfaces with external CTAO systems, retrieving service products from the Data Processing and Preservation System (DPPS) and sending analysis products and data quality reports to the Science User Support System (SUSS).

\begin{figure}[htp]
 \centering
 \makebox[\textwidth][c]{\includegraphics[width=0.99\textwidth]{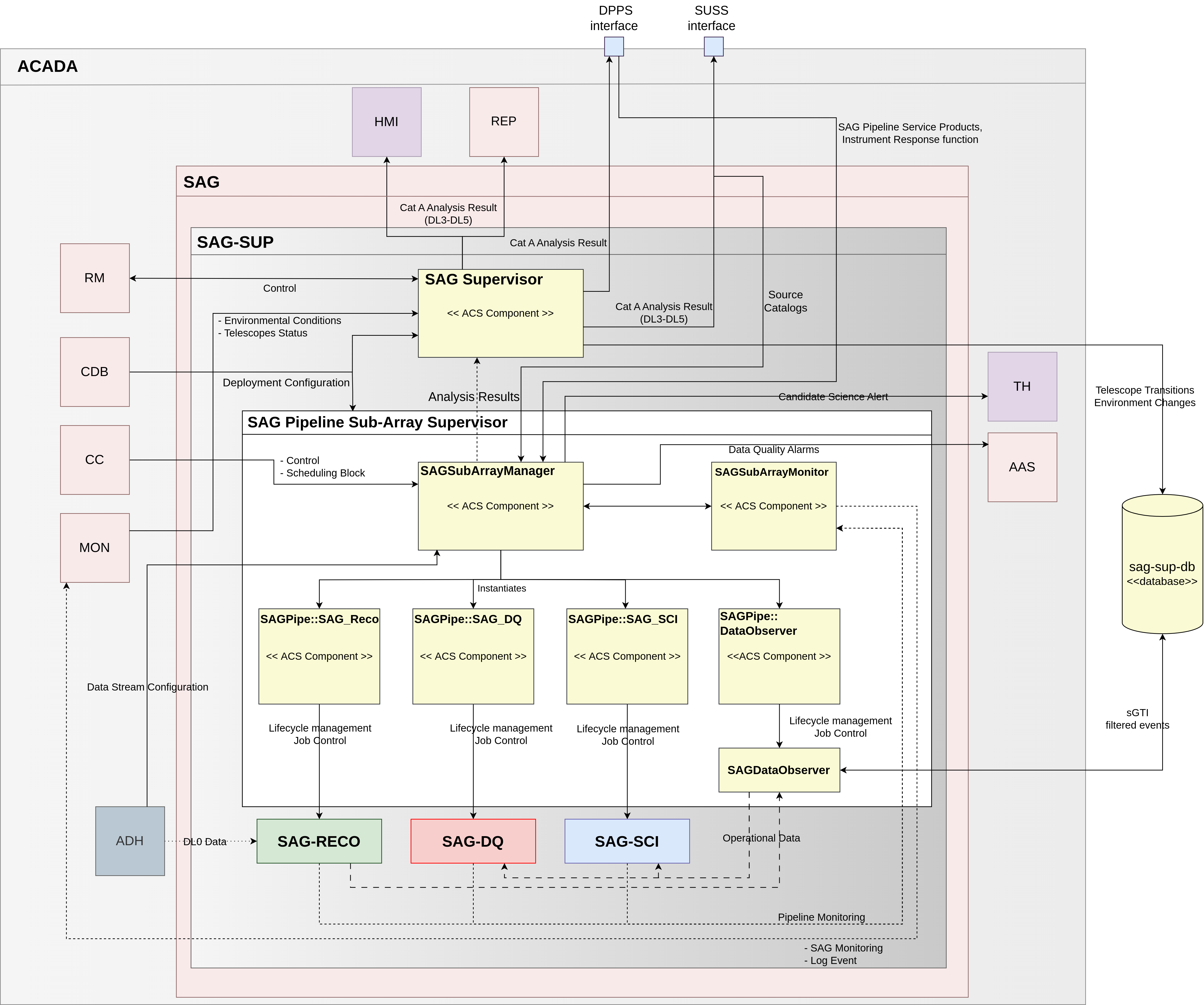}}
 \caption{
 Context diagram of the SAG-SUP package.
 Pink and purple boxes represent software components, which are ACADA sub-systems.
 The ports within the boundary of the ACADA box identify the system’s external interfaces.
 The dashed lines show the flow of data elements, and the arrows show the direction of the flow. SAG-SUP components are depicted with yellow boxes. The two main SAG-SUP components are the Supervisor, a unique component responsible for generating reports, and the SAG Pipeline Sub-Array Supervisor, which handles the life cycle of the pipeline to process data from a sub-array.
 Several SAG Pipeline Sub-Array Supervisors can be instantiated simultaneously to work with different sub-arrays.}
 \label{fig:SUP_architecture}
\end{figure}

The SAG pipeline is composed of three main components and a supervisor:
SAG-RECO performs low-level data reconstruction from DL0 to DL3 data in real-time, extracting event parameters (direction, energy, particle type) using machine learning models.
SAG-DQ executes online data quality checks from DL0 to DL3, identifying anomalies or degraded conditions through a configurable graph-based system.
SAG-SCI carries out high-level scientific analyses on DL3 data, detecting candidate transient events and issuing alerts accordingly.
SUP is the component of SAG interfacing the analysis pipelines with ACADA. The general view of the SAG-SUP in the ACADA context is shown in Fig.~\ref{fig:SUP_architecture}, where the interfaces with the other sub-systems of ACADA and with the pipelines are highlighted.
SAG-SUP is composed of two main components: \textit{SAG Supervisor} and \textit{SAG Pipeline Sub-array Supervisor}.

\paragraph{SAG Supervisor.}
SAG Supervisor is a unique ACS component instantiated by the RM with the same lifetime as ACADA.
It collects environmental conditions from the MON system and updates on telescope transitions described in this contribution (section~\ref{sec:newGTI}).
In addition, it subscribes to the CC notification channels to receive status changes and observation identifiers for ongoing observations and generates scientific monitoring results for the REP and HMI systems.

\paragraph{SAG Pipeline Sub-array Supervisor.}
The SAG Pipeline Sub-array Supervisor is responsible for managing the analysis of data from a sub-array of telescopes.
When multiple sub-arrays observe different targets simultaneously, their data can be processed in parallel and in real-time by separate Sub-array Supervisors.
A Sub-array Supervisor is instantiated by the CC at the beginning of the observations of a sub-array.
When observations end, the component ensures that the SAG pipelines are shut down, terminating data processing in a controlled way.
The Sub-array Supervisor comprises several ACS dynamic elements: the SAGSubArrayManager, the SAGSubArrayMonitor and four SAGPipe.
The SAGSubArrayManager is the main interface between the SAG system and ACADA during observations, managing requests from the CC and ADH for the sub-array observations and retrieving the deployment configuration from the CDB.
The SAGSubArrayManager is an event-driven component and possesses an internal finite state machine governing the permissible sequence of operations, ensuring the proper life cycle and response of the pipelines are respected.
The SAGSubArrayManager creates four SAGPipe components, three of which control SAG-RECO, SAG-DQ and SAG-SCI.
The SAGPipe elements configure the sub-systems and execute their processes, using the slurm Workload Manager to distribute the workload in the data centre.
The fourth SAGPipe, controlling the SAGDataObserver and the related database are discussed in section~\ref{sec:newGTI}. 
Finally, the SAGSubArrayManager creates a SAGSubArrayMonitor to monitor and collect operational data from the individual pipelines, such as processing rates.
These data are sent to the MON sub-systems through an ACS notification channel.

\section{Enhancing the Analysis of High-Priority Sources: Selecting Tracking Telescopes}
\label{sec:startDataTaking}
One of the main improvements of SAG-SUP to satisfy SAG requirements aims to handle multi-telescope operations and obtain the time when each telescope starts tracking.
In ACADA Rel1, SAG operates with a single telescope and starts analysing the data when the telescope ends the slewing phase and begins tracking the target.
In particular, the CC informs SAG-SUP with a dedicated StartOB() method, where OB stands for Observing Block, a unit of uninterrupted observations on a specific target.
Analysing data with the single telescope or a sub-array tracking the target allows us to select the set of Random Forests (RFs) Instrument Response Functions (IRFs) specific to the sub-array configuration.
The RFs are used by SAG-RECO to reconstruct the gamma-ray properties of Cherenkov events, while IRFs are used by SAG-SCI to estimate the flux of the observed sources.
Both of them are pre-computed offline using simulated data and sub-arrays with telescopes during tracking mode, excluding telescopes in slewing mode due to the complexity of the simulations and the small amount dedicated to slewing mode compared to the tracking mode.
For this reason, the reconstruction and analysis of events acquired by slewing telescopes is affected by large systematic errors, which significantly affect the data quality, and is not implemented by SAG.

The regular behaviour of SAG-SUP, which waits for all telescopes to be in tracking, will still be used in ACADA Rel2 for observations of regular sources, for which the delay between the times when the first and last telescopes start tracking is not significant.
The delay, which is due to different telescope types, repointing speeds and distances from the previous pointing position, may take up to a few tens of seconds in the worst-case scenario.
This behaviour is not optimal for the observations of high-priority sources, such as GRBs, which are often observed in response to a science alert.
It is crucial to begin the analysis of high-priority sources as soon as possible to detect possible prompt emission or determine stringent upper limits in case of a non-detection, thus maximising the scientific return of the CTAO in the context of time domain astrophysics.

For this reason, SAG will record when each telescope starts tracking the target and analyse in real-time their data to issue candidate Science Alerts as soon as possible.
The data of telescopes still in a slewing phase is flagged and discarded for the analysis.
The tracking starting instant of each telescope is provided to SAG-SUP by CC calling a method startDataTaking(TelescopeID) of the SAGSubArrayManager, just after the CC sends to a telescope the command to start the observation (tracking).
The method is called separately for each telescope, and when the SAGSubArrayManager receives the call, it updates the telescope status from TRACKING to OBSERVING registering a timestamp for the transaction.
We note that a small time delay exists between the CC command and its execution by the telescope hardware, so the tracking start timestamp registered by SAG-SUP is an approximation of the real value.

To realise this workflow, SAG-SUP uses telescope status timestamps to define \textbf{Good Time Intervals} (GTIs) and discard slewing data. A GTI is a time window during which a sub-array satisfies all conditions required for valid gamma-ray event acquisition, and events not belonging to a GTI window must be discarded from the analysis. As a result, larger GTI windows lead to an increase in exposure and sensitivity of the analysis.
As a result of the selection, SAG-SCI will analyse the SAG-RECO reconstructed events taken when at least one telescope was TRACKING to determine the statistical significance of the target source.
If the source is detected, SAG will issue a candidate science alert providing the significance.
The flux cannot be provided at this stage due to the absence of valid IRFs, and it can be provided in a later update when all the telescopes are in tracking mode.
Nevertheless, the estimation of the significance is sufficient to issue a candidate science alert, as the flux will be affected by real-time biases, such as the lack of knowledge of the redshift in the case of extragalactic objects like GRBs. We define two types of GTIs: standard and enhanced (section \ref{sec:newGTI}).

\textbf{Standard GTIs} include only information of the telescope status: a GTI can start when at least one telescope of the sub-array is in TRACKING mode and ends when none are tracking any more; additional configuration can define the minimal number of tracking telescopes to get a GTI. 
SAGSubArrayManager inserts the tracking timestamps sent by the CC described above in a newly introduced database (sag-sup-db in Fig.~\ref{fig:SUP_architecture}), which will record which telescopes are in a tracking status so they can be used from SAGDataObserver component described in the section \ref{sec:newGTI}, which builds the standard GTIs.

\section{Enhanced good time interval and SAGDataObserver }\label{sec:newGTI}
The second significant enhancement of SAG-SUP is to include the available monitoring and Data Quality information in the GTI determination to improve the data analysis and a component to determine GTIs and operate with them. \textbf{Enhanced GTI} extends the Standard GTI definition by applying additional selection criteria and depends on the availability of information from the MON system and Data Quality assessments. The MON system collects information from on-site hardware, including telescope statuses—such as structure, camera, and pointing direction—reported by the telescope managers. MON collects monitoring points and publishes them every few seconds via Apache Kafka,\footnote{\url{https://kafka.apache.org/}} a distributed event streaming platform designed for high-throughput, fault-tolerant data pipelines and real-time analytics; these data are then consumed by the SAG Supervisor ACS component (see Section~\ref{sec:SAGSUP}) through dedicated Kafka topics.
This information allows SAG-SUP to improve GTI reliability by, for example, identifying telescopes that become unavailable after the observation is configured and ensuring that TRACKING timestamps maximum error is one second. Environmental data collected via a separate Kafka topic include parameters such as humidity, cloud coverage, atmospheric transmission and extinction, and estimates of the Night Sky Background. These exclude time intervals where the conditions fall outside the predefined thresholds for a given observation. Enhanced GTIs are further constrained by the presence of vetoes from SAG-DQ, which flags periods of poor data quality based on DL1 and DL2 data quality analysis. Monitoring information and DQ-vetoes are stored in the sag-sup-db. 

\begin{figure}[htp]
 \centering
 \makebox[\textwidth][c]{\includegraphics[width=0.99\textwidth]{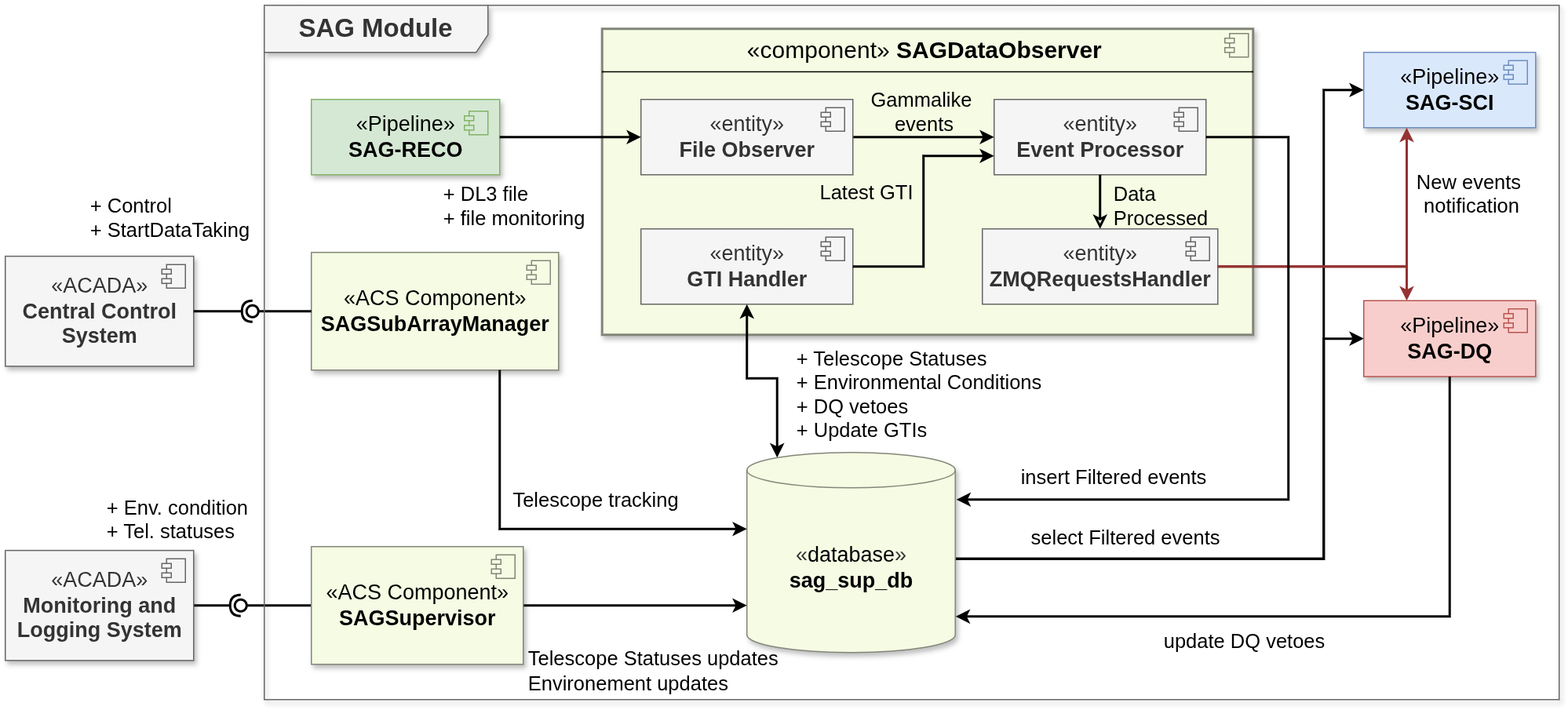}}
 \caption{Logical workflow of the SAGDataObserver component. SAG-RECO provides event files and file-monitoring. SAGSubArrayManager and SAGSupervisor supply telescope tracking timestamps and environmental condition updates, respectively. SAG-DQ generates data quality vetoes. SAGDataObserver aggregates all this information in the sag-sup-db database to generate Enhanced GTIs, filter event data accordingly, and provide reliable events to SAG-DQ and SAG-SCI pipelines. Notification messages inform the pipelines when new filtered events are available in the database.}
 \label{fig:SUP_dataobserver}
\end{figure}

The SAG-SUP element which actively creates and uses the GTIs is the SAGDataObserver, newly developed for ACADA Rel2 centralising the selection of the good events required from SAG-SCI and SAG-DQ DL3 analysis. SAGDataObserver acts as an event filter, ensuring that the SAG high-level pipelines process only reliable data while supporting the integration of an event-driven operational workflow. Any Science Alerts or analysis results associated with time windows not included in GTI are flagged or vetoed by SAG accordingly. The logical workflow of the SAGDataObserver is shown in Fig.~\ref{fig:SUP_dataobserver}.
SAGDataObserver has several logical components used to monitor the processing of data in SAG:
\begin{itemize}[noitemsep, leftmargin=.15in]
    \item \textbf{File Observer.}
    Monitors the production of files from SAG-RECO, which must then be analysed by SAG-DQ and SAG-SCI. Monitoring information is collected via ZeroMQ\footnote{\url{https://zeromq.org/}} (ZMQ) notification messages. ZMQ is an open-source messaging library.
    
    \item \textbf{GTI Handler.}
    Queries the sag-sup-db fetching the most updated Telescope information, monitoring information, and DQ vetoes, aggregating and computing the GTI every time the Supervisor and SubArrayManager components update the database. Standard and enhanced GTI are stored in different tables.
    
    \item \textbf{Event Processor.}
    Reads gamma-like reconstructed event files produced by SAG-RECO, filters events belonging to GTIs, and inserts them into the database to be used by SAG-SCI.
    
    \item \textbf{ZMQRequestsHandler.}
    Dispatches notification messages (ZMQ) to other SAG components to enable event-based processing. The dispatched information includes the presence of new filtered events with GTIs.
\end{itemize}

\section{Discussion and Conclusion}
\label{sec:conclusion}
We presented the enhanced version of SAG-SUP proposed to satisfy the SAG functional requirements.
The first significant improvement is the development of the startDataTaking method, which will allow SAG to know which telescopes are starting to track and immediately begin the high-level data analysis of high-priority sources such as GRBs.
This represents a significant advancement for CTAO's real-time science capabilities, as it will allow time-critical detections of transients and the issuing of candidate Science Alerts.

The second significant enhancement is the development of the SAGDataObserver, which will construct standard GTIs with the telescope status and enhanced GTIs, including environmental and telescope information from the MON system and data quality vetoes from SAG-DQ.
The knowledge or estimation of such information is used by SAG to select good-quality events, which is a mission-critical factor that directly impacts the overall quality of the results and the reliability of each science alert issued.

The architecture of the SAG pipeline is thus evolving based on the experience gained over the years and the commitment to achieving optimal results in real-time processing, enhancing the interconnectivity of pipelines and communication within the ACADA system.
The SAG pipeline will allow the CTAO to become a crucial instrument for time-domain astrophysics and enable multi-wavelength and multi-messenger approaches that will lead to a deeper understanding of the broad-band non-thermal properties of target sources.

\section*{Acknowledgements}
\noindent We gratefully acknowledge financial support from the agencies and organisations listed here:

\noindent \href{https://www.ctao.org/for-scientists/library/acknowledgments/}{https://www.ctao.org/for-scientists/library/acknowledgments/}


\end{document}